\documentclass[fleqn,usenatbib]{mnras}

\usepackage{newtxtext,newtxmath}

\usepackage[T1]{fontenc}

\DeclareRobustCommand{\VAN}[3]{#2}
\let\VANthebibliography\thebibliography
\def\thebibliography{\DeclareRobustCommand{\VAN}[3]{##3}\VANthebibliography}

\newcommand{\dk}[1]{{\textcolor{black}{#1}}}

\usepackage{graphicx}	
\usepackage{amsmath}	

\title[Black Hole Envelope]{Black Hole Envelopes in Little Red Dots}

\author[D. Kido et al.]{
Daisaburo Kido,$^{1}$\thanks{E-mail: dkido@g.ecc.u-tokyo.ac.jp}
Kunihito Ioka,$^{2}$
Kenta Hotokezaka,$^{1}$
Kohei Inayoshi,$^{3}$
and Christopher M. Irwin$^{1}$
\\
$^{1}$ Research Center for the Early Universe, Graduate School of Science, The University of Tokyo, Bunkyo, Tokyo\\
$^{2}$ Yukawa Institute for Theoretical Physics, Kyoto University, Kyoto 606-8502, Japan\\
$^{3}$Kavli Institute for Astronomy and Astrophysics, Peking University, Beijing 100871, China\\
}

\date{Accepted XXX. Received YYY; in original form ZZZ}

\pubyear{\the\year{}}

\begin{document}
\label{firstpage}
\pagerange{\pageref{firstpage}--\pageref{lastpage}}
\maketitle

\begin{abstract}
Recent observations by the James Webb Space Telescope have uncovered a population of compact, red object ($z\sim 4\text{--}7$) known as little red dots (LRDs). The presence of broad Balmer emission lines \dk{implies} active galactic nuclei powered by supermassive black holes (BHs), while LRDs exhibit unusually weak X-ray and radio emission and low variability, suggesting super-Eddington accretion that obscures the central engine. 
We suggest that such an extreme accretion disc inevitably drives strong outflows, which would disrupt the LRDs themselves unless confined within the nuclear region -- posing a general feedback problem for overmassive BHs. 
To resolve this, we propose that the BH is embedded in a massive, optically thick envelope that gravitationally confines the outflow, making any outflow a no-go.
This envelope, powered by accretion on to the BH, radiates at nearly the Eddington limit, and is sustained by an infall of the interstellar medium at a rate on the order of $\sim 1 M_{\odot}~{\rm yr}^{-1}$.
A photosphere emerges either within the envelope or in the infalling medium, with a characteristic temperature of $5000$–$7000~\mathrm{K}$, near the Hayashi limit. The resulting blackbody emission naturally explains the red optical continuum of the distinct V-shaped spectrum observed in most LRDs. Furthermore, the dynamical time-scale at the photosphere, $\sim 0.01~{\rm pc}$, is consistent with the observed year-scale variabilities. The nuclear structure and spectral features of LRDs are shaped by this envelope, which not only regulates feedback but also acts as a gas reservoir that sustains rapid BH growth in the early universe.
\end{abstract}

\begin{keywords}
galaxies: high-redshift  -- quasars: supermassive black holes  -- galaxies: active  
\end{keywords}



 \section{Introduction}
\label{sec:introduction}
The James Webb Space Telescope (JWST) has 
discovered a new population of red and compact objects in the high-redshift universe, so-called ``Little Red Dots'' 
\citep[LRDs,][]{Harikane+:2023, Kocevski2023ApJ, Kocevski2025ApJ, Akins2024arXiv, Matthee2024ApJ, Labbe2025ApJ}.
Among the most intriguing spectral features of LRDs is a characteristic ``V-shaped'' transition from a flat blue to a steep red continuum 
in the spectral energy distributions (SEDs) across rest-frame UV to optical wavelengths \dk{with the inflection occuring near the Balmer limit \citep{Setton2024arXiv}. }
LRDs, selected by proper colour selection criteria, exhibit broad Balmer emission lines with full width half maximum (FWHM) of $\gtrsim 1000~{\rm km\,s^{-1}}$, 
suggesting that LRDs are active galactic nuclei (AGNs) powered by massive black holes (BHs) with masses of $\sim 10^{6-8}~M_\odot$
\citep{Matthee2024ApJ,Greene+:2024,Kocevski2025ApJ,Taylor2025ApJ, Kokorev+:2024}.
If one instead interprets these V-shaped SEDs as arising from massive galaxies \dk{\citep{Labbe2023Natur}}, \dk{the broad Balmer emission may be explained by the velocity dispersions of the gas \citep{Baggen2024ApJ}}, but the implied stellar masses and dust contents would be in tension with the current $\Lambda\mathrm{CDM}$ model prediction \citep{Boylan-Kolchin2023NatAs}. 
The LRD population is not rare but rather abundant with cosmic number densities of $\Phi \sim 10^{-5}- 10^{-4}~\mathrm{Mpc^{-3}}$ \dk{\citep{Harikane+:2023, Matthee2024ApJ, Greene+:2024, Kocevski2025ApJ, Kokorev+:2024}},
which is $\sim 1-2$ orders of magnitude higher than that of UV-bright quasars \citep[e.g.,][]{McGreer2018AJ, Matsuoka2018ApJ,Niida2020ApJ,Matsuoka2023ApJ}.

Observational studies of LRDs have unveiled that 
their properties differ
from typical AGNs.
One of the notable characteristics of LRDs is their unusually weak X‑ray emission. Although multi‑wavelength strategies typically include X‑ray and radio observations to detect AGNs that are heavily obscured in the UV/optical bands, recent studies have shown that JWST‑identified AGNs remain undetected in X‑rays \citep{Kocevski2023ApJ, Yue2024ApJ, Juodzbalis2024MNRAS, Ananna2024ApJ, Maiolino2025MNRAS, Akins2024arXiv}.
Similarly, radio observations—often employed to trace jet activity or other radio-loud features—have not revealed the expected signatures of an AGN in these systems \citep{Mazzolari2024arXiv, Gloudemans2025ApJ, Perger2025A&A}.
This simultaneous lack of X‑ray and radio emission suggests that the central engines in LRDs may either be intrinsically faint in these bands \citep{Inayoshi+KN:2024} or that their emission is being suppressed or absorbed by surrounding material \citep{Maiolino2025MNRAS}.
Furthermore, observations in the infrared offer additional insight into the emission from hot dust in the AGN torus, which is typically prominent. However, several studies show that the flux density in the JWST MIRI bands is considerably fainter than the value expected from the ordinary AGN dust torus model \citep{Williams2024ApJ, Akins2024arXiv, Setton2025arXiv}. Although a subset of LRDs exhibits relatively strong infrared flux \citep{PerezGonzalez2024ApJ, Lyu2024ApJ, Barro2024arXiv}, such behaviour appears to characterize a distinct class of objects that deviates from the expectations of the conventional AGN dust torus model. 
Besides, emission from AGN accretion discs is known to exhibit a temporal variability on time-scales ranging from hours to years \citep{Ulrich1997, Kelly2009ApJ, MacLeod2010ApJ, MacLeod2012ApJ}. 
\dk{Recent observations with JWST and HST in rest-frame UV and optical has not revealed clear evidence for significant variability over limited monitoring epochs (typically two or three visits; \citealt{Kokubo2024arXiv, Zhang2025ApJ, Tee2025ApJ}), while a subset of LRDs, $\sim 3\%$ of all photometric LRD candidates, show detectable variability on time-scales of years \citep{Zhang2025ApJ, Ji2025arXiv, Furtak2025A&A}. 
Even after several years of JWST operation, clear rapid fluctuations have yet to be observed, adding further challenges to the dust-obscured AGN interpretation of the LRDs.}

The spectral properties of LRDs compared to AGNs -- namely the absence of detectable X-ray and radio emission -- lead to the hypothesis that they are powered by super-Eddington accretion, which becomes optically thick near the BH \dk{\citep{Pacucci2024ApJ,Madau2024ApJ,Lambrides2024arXiv,Inayoshi+KN:2024, King2025MNRAS}}. 
However, super-Eddington discs are expected to drive powerful outflows\footnote{
The outflow may include jets launched by the Blandford-Znajek mechanism \citep{Blandford+Z:1977,McKinney+TSN:2014}.
Here we focus on winds, which are relatively independent of the BH spin and magnetic field.
(But, we reserve the term 'wind' for those originating from the BH envelope.)}
and strong feedback
\citep{Shakura+S:1973,Blandford+B:1999,Blandford+B:2004,Ohsuga+MNM:2005,Ohsuga+M:2011,Jiang+SD:2014,Sadowski+NMT:2014,Sadowski+:2015,Inayoshi2016MNRAS},
which should leave observable X-ray and radio signatures. 
This apparent contradiction raises a key question: under what conditions can this scenario be naturally explained, and when does it break down? 
In what follows, we examine these potential inconsistencies and explore the true nature of LRD nuclei.
In particular, we propose a scenario in which the confinement of the strong outflow is naturally achieved through the presence of a dense, extended envelope surrounding the BH.
A similar envelope structure has been discussed in the context of Eddington envelopes during tidal disruption events (TDEs) \citep{Loeb1997ApJ,Ulmer1998A&A,Metzger2022ApJ,Price+:2024}
and quasi-stars \citep{Begelman2008MNRAS,Ball+TZE:2011,Coughlin2024ApJ}
\dk{(with historical references going back to 
\cite{Landau:1938}, \cite{Berezinskii1977}, \cite{Thorne+Z:1977}, and \cite{Berezinskii1981MNRAS}).}
Here, we propose that the BH envelope could be essential in the case of LRDs to prevent the feedback. 
\footnote{
\dk{We refer to it as a BH envelope when the mass of the envelope is smaller than that of the BH. Conversely, when the envelope mass is larger and gas self-gravity affects the structure, we distinguish it as a quasi-star. As we will see later, unlike quasi-stars, the BH envelope solution is independent of the inner boundary conditions.
}}

\dk{In the proposed BH envelope framework, the dense gaseous envelope not only provides the necessary conditions for outflow confinement but also imprints distinct spectral signatures.}
One striking prediction is that the photospheric temperature of the envelope settles around $5000$-$7000~\mathrm{K}$. 
This temperature leads to red continua in the rest-frame optical, a feature that emerges naturally from the envelope’s structure without requiring additional attenuation.  
While this work was in preparation, \citet{Naidu2025arXiv, Rusakov2025arXiv, deGraaff2025arXiv} independently showed that dense material around the BH can give rise to a similar signature.
Our proposal goes one step further and makes a stronger claim.
The combination of outflow confinement by dense surrounding material and the resulting red spectral continuum offers a unified explanation for several puzzling observations of active galaxies in extreme accretion regimes.
Moreover, we reveal a connection between LRDs and the classical Hayashi track, in which the envelope’s surface temperature settles at the onset of hydrogen recombination \citep{Hayashi1961PASJ, HayashiHoshi1961PASJ}.
In this paper, we study the need for a dense envelope that engulfs the central BH so that strong momentum and energy feedback from the outflow does not affect the outer LRD.
Although we consider a simple model that assumes spherical symmetry and stationary conditions, and the structure may not follow our idealized treatment, it none the less provides insight into the envelope's structure and the associated observational signature.

This paper is organized as follows.
We present the feedback issues arising from super‐Eddington accretion in galaxies hosting massive BHs in Section \ref{sec:feedback}. 
In Section \ref{sec:envelope}, we introduce the BH envelope and outline its fundamental properties.
We then describe the detailed structure of the BH envelope and its observational signatures in section \ref{sec:structure}. 
Our conclusions are summarized in section \ref{sec:conclusions}.
Appendix \ref{ap:st} provides further details on the numerical procedures employed in Section \ref{sec:structure}, and Appendix \ref{ap:not} tabulates the notation used throughout the paper. 

We define the “LRD nucleus” as the region within the radius of the broad-line region, $r < r_{\mathrm{nuc}} \sim 3 \times 10^{16}~\mathrm{cm}$, and we refer to the volume within the typical LRD radius $r_\mathrm{LRD} \sim 100~\mathrm{pc}$ as the “entire LRD.” 
\dk{
The entire LRD includes not only the envelope, but also the reservoir of accreting gas outside the envelope, which may extend to significantly larger radii.
}
\section{Apparent Lack of Super-Eddington Feedback}
\label{sec:feedback}

Super-Eddington accretion discs are optically thick near the BH and generally produce radiation-driven outflows 
\citep{Shakura+S:1973,Blandford+B:1999,Blandford+B:2004},
with convective energy transport \citep{Narayan+IA:2000,Quataert+G:2000},
as demonstrated by various hydrodynamical simulations
\citep{Ohsuga+MNM:2005,Ohsuga+M:2011,Jiang+SD:2014,Sadowski+NMT:2014,Yuan+N:2014,Sadowski+:2015,Sadowski+N:2016,Kitaki+MOK:2018,Yoshioka+MOKK:2022,Hu+IHQK:2022,Toyouchi+HI:2024,Shibata2025PhRvD}.
There are also observational indications that faster outflows tend to be driven in higher-Eddington BHs \citep{Gofford+:2015,Matzeu+:2017,Laha+:2021}.
Due to the outflow, the mass accretion rate decreases with radius roughly as $\dot{M} \propto r^{0.6}$
\citep{Hu+IHQK:2022,Toyouchi+HI:2024,Shibata2025PhRvD},
and the mass accretion rate on to the BH ($\dot{M}_{\rm BH}$) is only a fraction of the initial one.
The total mass outflow rate ($\dot{M}_{\rm out}$) exceeds $\dot{M}_{\rm BH}$, increasing with disc radius, while the energy generation takes place predominantly near the inner radius.
The total luminosity released from the disc is proportional to the accretion rate on to the BH,
\begin{align}
  L=\eta \dot{M}_{\rm BH} c^2,
\end{align}
where the radiative efficiency $\eta$ is 
set to $\eta=0.1$ as predicted in the standard thin-disc model \citep{Shakura+S:1973}.
The Eddington luminosity,
$L_{\rm Edd}=4\pi G c M_{\rm BH}/\kappa_{\rm T} \sim 1.2 \times 10^{45} (M_{\rm BH}/10^{7} M_{\odot})~\mathrm{erg~s^{-1}}$,
where $\kappa_{\rm T}=0.4~\mathrm{cm^{2}~g^{-1}}$ is the Thomson opacity,
is achieved for an Eddington accretion rate defined by
\begin{align}
  \dot{M}_{\rm Edd}=\frac{1}{\eta}\frac{L_{\rm Edd}}{c^{2}}
  =\frac{1}{\eta}\frac{4\pi G M_{\rm BH}}{c \kappa_{\rm T}}
  \sim 0.2 \left(\frac{M_{\rm BH}}{10^{7} M_{\odot}}\right) M_{\odot}~\mathrm{yr^{-1}}.
  \label{eq:dotMedd}
\end{align}
We normalize the BH accretion rate according to
\begin{align}
  \dot{m}\equiv \frac{\dot{M}_{\rm BH}}{\dot{M}_{\rm Edd}}.
  \label{eq:dotm}
\end{align}

\subsection{Momentum feedback}

First, we show that the momentum feedback of the disc outflow is powerful enough to disrupt the entire LRD.
Here, we focus on momentum rather than energy to make the argument independent of cooling processes.
The momentum outflow rate is proportional to the BH accretion rate $\dot{M}_{\rm BH}$, as the mass outflow rate scales with $\dot{M}_{\rm BH}$,
\begin{align}
  \dot{P}_{\rm out} = \xi c \dot{M}_{\rm BH}
  \sim 5 
  \left(\frac{\xi}{0.03}\right)
  \left(\frac{L}{10^{45}~\mathrm{erg~s^{-1}}}\right)
  M_{\odot}~\mathrm{cm~s^{-2}},
  \label{eq:dotPout}
\end{align}
where $\xi$ is the outflow efficiency
and we adopt $\xi = 0.03$ according to \cite{Hu+IHQK:2022}.
We emphasize that this is a conservative value.
If the outflow traps radiation and becomes adiabatic, i.e., if ${\dot M}_{\rm BH} \gg {\dot M}_{\rm Edd}$,
the outflow efficiency increases by an order of magnitude, reaching $\xi \gtrsim 0.4$,
because all the energy is transferred to the kinetic energy of the outflow $\dot{M}_{\rm out} v_{\rm out}^{2}/2 \sim \eta \dot{M}_{\rm BH} c^{2}$ and 
the total outflow rate exceeds the BH accretion rate $\dot{M}_\mathrm{BH} < \dot{M}_\mathrm{out}$,
which yield the momentum outflow rate
$\dot{P}_{\rm out} \sim \dot{M}_{\rm out} v_{\rm out} > \sqrt{2\eta} c \dot{M}_{\rm BH} \sim  0.4 c \dot{M}_{\rm BH}$. 
Additionally, radiation emitted from outflowing gas, especially in the UV, is effectively reabsorbed, contributing to the outflow momentum.

The total duration of the outflow is a fraction of the cosmic time $t_{\rm age} \sim 10^{9}$ yr at $z=4$--$7$,
\begin{align}
  t_{\rm out} \sim f_{\rm duty} t_{\rm age} \sim 3 \times 10^{7} \left(\frac{f_{\rm duty}}{0.03}\right)~\mathrm{yr},
  \label{eq:tout}
\end{align}
where $f_{\rm duty}$ is the duty cycle.
We adopt $f_{\rm duty} \sim 0.03$ as suggested by
the number density of LRDs $\Phi \sim 10^{-4.5}~\mathrm{cMpc^{-3}~mag^{-1}}$,
which constitutes $\sim 3\%$ of the galaxy population at
$M_{\rm UV} \sim -19$ and $z=4$--$7$ \citep[e.g.,][]{Harikane+:2023,Kocevski2025ApJ,Kokorev+:2024}.
Note that $f_\mathrm{duty} = 0.03$ is regarded as a lower bound, since it assumes every galaxy undergoes an LRD phase. 
This duration $t_{\rm out} \sim 3 \times 10^{7}~\mathrm{yr}$ is also comparable to the Salpeter time
\begin{align}
  t_{\rm Sal}=\frac{M_{\rm BH}}{\dot{M}_{\rm Edd}} \sim 4.5 \times 10^{7}~\mathrm{yr}.
  \label{eq:tSal}
\end{align}
Using equations (\ref{eq:dotPout}) and (\ref{eq:tout}),
the total outflow momentum ($P_{\rm out} = t_{\rm out} \dot{P}_{\rm out}$) is estimated as
\begin{align}
  P_{\rm out} 
  \sim  5 \times 10^{15}
  \left(\frac{f_{\rm duty}}{0.03}\right)
  \left(\frac{\xi}{0.03}\right)
  \left(\frac{L}{10^{45}~\mathrm{erg~s^{-1}}}\right)
  ~M_{\odot}~\mathrm{cm~s^{-1}}.
  \label{eq:Pout}
\end{align}

\begin{figure*}
    \centering
    \includegraphics[width=0.9\linewidth]{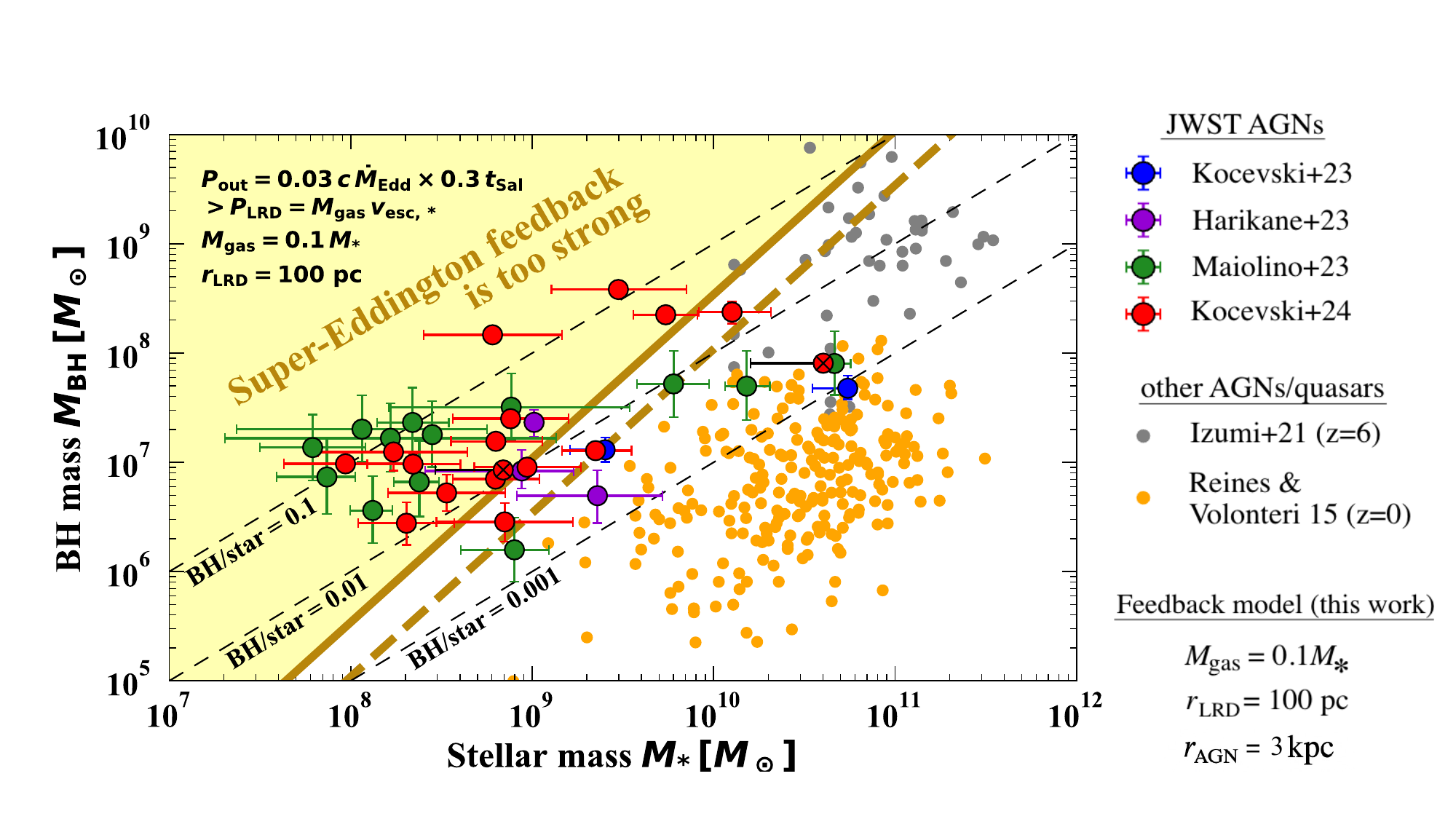}
    \caption{
      The BH mass $M_{\rm BH}$ versus stellar mass $M_{*}$ relation compared with
      the too-strong-feedback condition (gold solid line) in equation (\ref{eq:feedback})
      using the outflow efficiency $\xi=0.03$ in equation (\ref{eq:dotPout}),
      the ratio of the outflow time to the Salpeter time $f_{\rm Sal}=t_{\rm out}/t_{\rm Sal}=0.3$,
      the ratio of the gas mass to the stellar mass $\zeta_{\rm gas}=M_{\rm gas}/M_{*}=0.1$,
      the Eddington ratio $\dot{m}=1$
      and the LRD radius $r_{\rm LRD}=100~\mathrm{pc}$. The gold dashed line indicates the same criterion adopting $r_\mathrm{AGN} = 3\,\mathrm{kpc}.$
      The dashed black lines indicate loci of constant BH to stellar mass ratio, $M_\mathrm{BH} / M_* = 0.001, 0.01, 0.1$ respectively.
      A large fraction of LRDs would be expected to be strongly affected by the outflow if such an outflow from a super-Eddington disc existed.
      The data are taken from \citet{Kocevski2025ApJ} and based on \citet{Kocevski2023ApJ,Reines+V:2015,Izumi+:2021,Harikane+:2023,Maiolino2024A&A}. 
      \dk{Note that these BH masses are estimated with the local scaling relations based on Balmer emission line FWHM and luminosity provided by \citet{Reines+V:2015, Greene2005ApJ}.}
      The two X-ray detected LRDs are marked with an X \citep{Kocevski2025ApJ}.
    }
    \label{fig:feedback}
\end{figure*}

On the other hand, the momentum required to disrupt the entire LRD is determined by the size and mass contained in the interior.
The typical radius of observed LRDs is approximately $r_{\rm LRD} \sim 100~\mathrm{pc}$ \citep{Baggen+:2023,Furtak+:2023,Guia+PK:2024}.
The mass of stellar components in the entire LRD depends on the SED fitting procedure.
If a stellar component is responsible for the rest-UV spectral part of the V-shaped SED, 
the stellar mass is required to be $M_{*} \sim 10^{9} M_{\odot}$
\citep[e.g.,][]{Wang2025ApJ,Kocevski2025ApJ}.
\dk{The gas mass can be expressed as
\begin{align}
  M_{\rm gas} = \zeta_\mathrm{gas} M_* = 10^8 \left( \frac{\zeta_\mathrm{gas}}{0.1} \right) \left(\frac{M_*}{10^9 M_\odot} \right)M_\odot,
  \label{eq:Mgas}
\end{align}
where $\zeta_\mathrm{gas} = M_\mathrm{ gas}/M_*$ is the gas-to-stellar mass ratio, which is highly uncertain.  
In our fiducial model, we adopt $\zeta_\mathrm{gas} = 0.1$, corresponding to $M_\mathrm{gas} \sim 10^8\,M_\odot$.  
This value is consistent with the hydrogen column density of $N_\mathrm{H} \sim 10^{23}\,\mathrm{cm^{-2}}$ inferred from X-ray spectral analyses of a few LRDs \citep{Kocevski2025ApJ}, assuming the absorption mainly occurs at $r_{\rm LRD} \sim 100~\mathrm{pc}$, $M_\mathrm{gas} \sim 4 \pi r_\mathrm{LRD}^2 N_\mathrm{H} m_\mathrm{p} \sim 10^8 M_\odot.$
}

The escape velocity from LRDs is $v_{\rm esc, *} \sim (2 GM_{*}/r_{\rm LRD})^{1/2} \sim 300 (M_{*}/10^{9}M_{\odot})^{1/2} (r_{\rm LRD}/100~{\rm pc})^{-1/2}~\mathrm{km~s^{-1}}$,
where we neglect the gravitational potential of dark matter mass at $r\lesssim r_{\rm LRD}$.
Thus, since the BH outflow sweeps up only the gas component, one can calculate the momentum required for the surrounding gas to be unbound as
$P_{\rm LRD} = M_{\rm gas} v_{\rm esc, *}$ or 
\begin{align}
  P_{\rm LRD} 
  &\sim 3 \times 10^{15}
  \left(\frac{M_{\rm gas}}{10^{8} M_{\odot}}\right)
  \left(\frac{M_{*}}{10^{9} M_{\odot}}\right)^{1/2}
  \left(\frac{r_{\rm LRD}}{100~\mathrm{pc}}\right)^{-1/2}
  M_{\odot}~\mathrm{cm~s^{-1}}.
  \label{eq:PLRD}
\end{align}

Comparing equations (\ref{eq:Pout}) and (\ref{eq:PLRD}), the outflow momentum can exceed the momentum required for unbinding LRDs,
\begin{align}
  P_{\rm out} \gtrsim P_{\rm LRD}.
  \label{eq:Pout>PLRD}
\end{align}
Therefore, the BH outflow inevitably affects the LRD and may even completely destroy it in some cases 
(e.g., $\xi \gtrsim 0.03$ or $M_{*} \lesssim 10^{9} M_{\odot}$).
If the outflow is bipolar as expected and sweeps only part of the gas, the condition is more easily satisfied.
The impact is more severe in the inner region.
For example, around the broad-line regions located at $r_{\rm BLR} \sim GM_{\rm BH}/\Delta v^{2} \sim 3\times 10^{16} (M_{\rm BH}/10^{7} M_{\odot})~\mathrm{cm}$ for a line width $\Delta v \sim 2000~\mathrm{km~s^{-1}}$,
the gas mass is only $M_{\rm gas}\sim 4\pi r_{\rm BLR}^{2} N_{\rm H} m_{\rm p} \sim 20 M_{\odot}$ and much smaller than equation (\ref{eq:Mgas}).
\dk{We base our stellar mass estimates on SED modeling under the dust‐obscured AGN model assumption. If, however, the Balmer break arises from obscuring gas rather than starlight \citep{Inayoshi+M:2025}, these SED‐based masses can be drastically overestimated \citep{Wang2024ApJ}.  Indeed, clustering and halo mass analyses indicate typical LRD stellar masses of the order of  $10^7M_\odot $ \citep{Matthee2025ApJ, Carranza-Escudero2025arXiv}. However, according to equation (\ref{eq:PLRD}), it becomes easier to blow off the gas component if the stellar mass is smaller. Therefore, we conservatively adopt $M_*\sim 10^9 M_\odot $ to make our statement more robust.} 

The too-strong-feedback condition in equation (\ref{eq:Pout>PLRD})
can be rewritten as a relation between the BH mass $M_{\rm BH}$ and stellar mass $M_{*}$ by using equations (\ref{eq:dotm}), (\ref{eq:dotPout}), (\ref{eq:tSal}), (\ref{eq:Pout}), (\ref{eq:Mgas}) and (\ref{eq:PLRD}),
\begin{align}
  M_{\rm BH} > \frac{\zeta_{\rm gas}}{\xi f_{\rm Sal} \dot{m}} M_{*} \left(\frac{2GM_{*}}{r_{\rm LRD} c^{2}}\right)^{1/2},
  \label{eq:feedback}
\end{align}
where $f_{\rm Sal}=t_{\rm out}/t_{\rm Sal}$ is the ratio of the outflow duration to the Salpeter time.
Figure~\ref{fig:feedback} compares this relation with observations \citep{Reines+V:2015, Izumi+:2021, Kocevski2023ApJ, Harikane+:2023, Maiolino2024A&A, Kocevski2025ApJ}
using $\xi=0.03$, $f_{\rm Sal}=0.3$, $\zeta_{\rm gas}=0.1$, $\dot{m}=1$ and $r_{\rm LRD}=100~\mathrm{pc}$.
We can see that a large fraction of LRDs would be strongly affected by the feedback if the outflow from the super-Eddington disc were to reach the observed region. 
It is intriguing that LRDs with detected X-ray emission (marked with an X) do not satisfy this condition, or are just at the threshold.

\subsection{Energy feedback}

Next, let us consider the energy feedback.
The outflow luminosity associated with equation (\ref{eq:Pout})
is approximately $L_{\rm out} \sim 10^{43} (L/10^{45}~\mathrm{erg~s^{-1}})~\mathrm{erg~s^{-1}}$, which is again a conservative estimate (only $1\%$ of $L$).
The total energy released during the active time $t_{\rm out}$ in equation (\ref{eq:tout}) is $\sim 10^{58} (L/10^{45}\mathrm{erg~s^{-1}})~\mathrm{erg}$,
which is much larger than the binding energy of the gas contained in LRDs
$\sim GM_{*}M_\mathrm{gas}/r_{\rm LRD} \sim 10^{56}~\mathrm{erg}$. 
Therefore, if cooling is ineffective, the BH outflow will unbind the entire LRD.

If cooling is effective, the injected energy will primarily be released in the X-ray bands.
This is because the BH outflow velocity would exceed $v_{\rm out} \gtrsim 3000~\mathrm{km~s^{-1}}$ \citep{Hu+IHQK:2022,Shibata2025PhRvD}
and the initial temperature of the shock is $kT \sim m_{\rm p} v_{\rm out}^{2} \gtrsim 100~\mathrm{keV}$.
At this temperature, free-free emission dominates \citep{Sutherland+D:1993}
and most of the emission would occur in the X-ray regime.
Similarly to stellar-wind bubbles and supernova remnants, there are two shocks:
a forward shock propagating into the interstellar gas
and a reverse shock or termination shock within the outflow.
At least one of the shocks is hot and emits X-rays.
As the shock moves beyond the broad-line region
$r_{\rm{BLR}} \sim 10^{16} (M_{\rm BH}/10^{7}M_{\odot})~\mathrm{cm}$
(which occurs quickly, within a year),
the column density of 
$N_{\rm H} \sim 10^{23}~\mathrm{cm^{-2}}$ is insufficient to obscure X-rays.
The expected X-ray luminosity is
$L_{\rm X} \sim L_{\rm out} \sim 10^{43} (L/10^{45}~\mathrm{erg~s^{-1}})~\mathrm{erg~s^{-1}}$,
which is likely to lead to tension with
X-ray upper limits
\citep{Kocevski2023ApJ,Yue2024ApJ,Juodzbalis2024MNRAS,Maiolino2025MNRAS,Akins2024arXiv,Inayoshi+KN:2024}.

In conclusion, when the outflow reaches observable regions,
various tensions will likely arise, suggesting that outflows are a ``no-go.''
These feedback problems arise mainly because the BH mass is unusually large compared to the galaxy mass (see Figure~\ref{fig:feedback}). In other words, this is a common issue in models with overmassive BHs \citep{Inayoshi+:2022,Hu+:2022}.
Note that simply reducing the BH mass while keeping the luminosity fixed does not solve the issue, as it only increases the super-Eddington ratio $\dot{m}$ (and in fact makes things even worse by increasing $\xi$).

\section{AGN BH envelope}
\label{sec:envelope}

\begin{figure}
    \centering
    \includegraphics[width=1\linewidth]{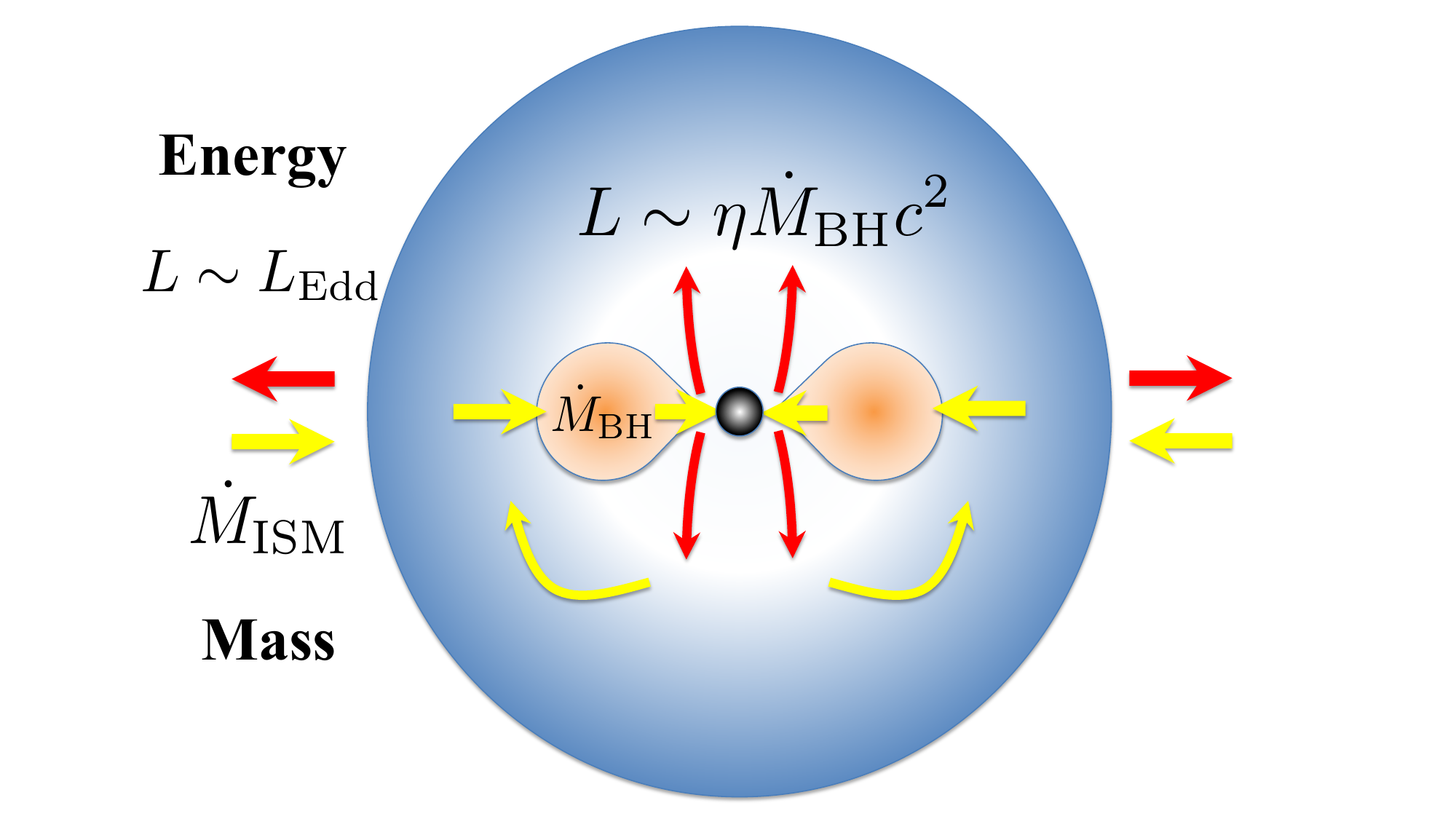}
    \caption{Concept of the BH envelope.
      The BH envelope serves to block and gravitationally confine the outflow
      from the super-Eddington disc.
      Meanwhile, energy is transported through radiation and convection, and
      radiated away from the envelope surface.
      The luminosity is primarily determined by the mass accretion rate
      on to the BH,
      $L \sim \eta \dot{M}_{\rm BH} c^{2}$,
      and is assumed here to be regulated to the Eddington luminosity
      of the entire system.
      Mass inflow from the interstellar medium (ISM) into the envelope
      $\dot{M}_{\rm ISM}$ is also considered.}
    \label{fig:Envelope_schematic}
\end{figure}

One way to avoid the feedback problems discussed in Sec.~\ref{sec:feedback} is to prevent the outflow from emerging.
This can be achieved by placing an envelope around the BH,
which can block the outflow and confine it with the BH’s gravity.
The energy of the outflow is reprocessed by the envelope,
transferred via radiation and convection,
and eventually emitted from its surface like a star (but powered by the BH; 
see Figure~\ref{fig:Envelope_schematic}).
The outflow gas halted by the envelope is mixed with the envelope material through convection, and part of it falls back into the BH.
The gas that falls from the envelope on to the BH forms a disc around it. While a fraction of this gas accretes on to the BH, most of it is launched back as an outflow due to the super-Eddington accretion, likely resulting in the formation of a convective core.
Roughly speaking, the system can be interpreted as a BH surrounded by an inflated accretion disc.

\begin{figure}
    \centering
    \includegraphics[width=1\linewidth]{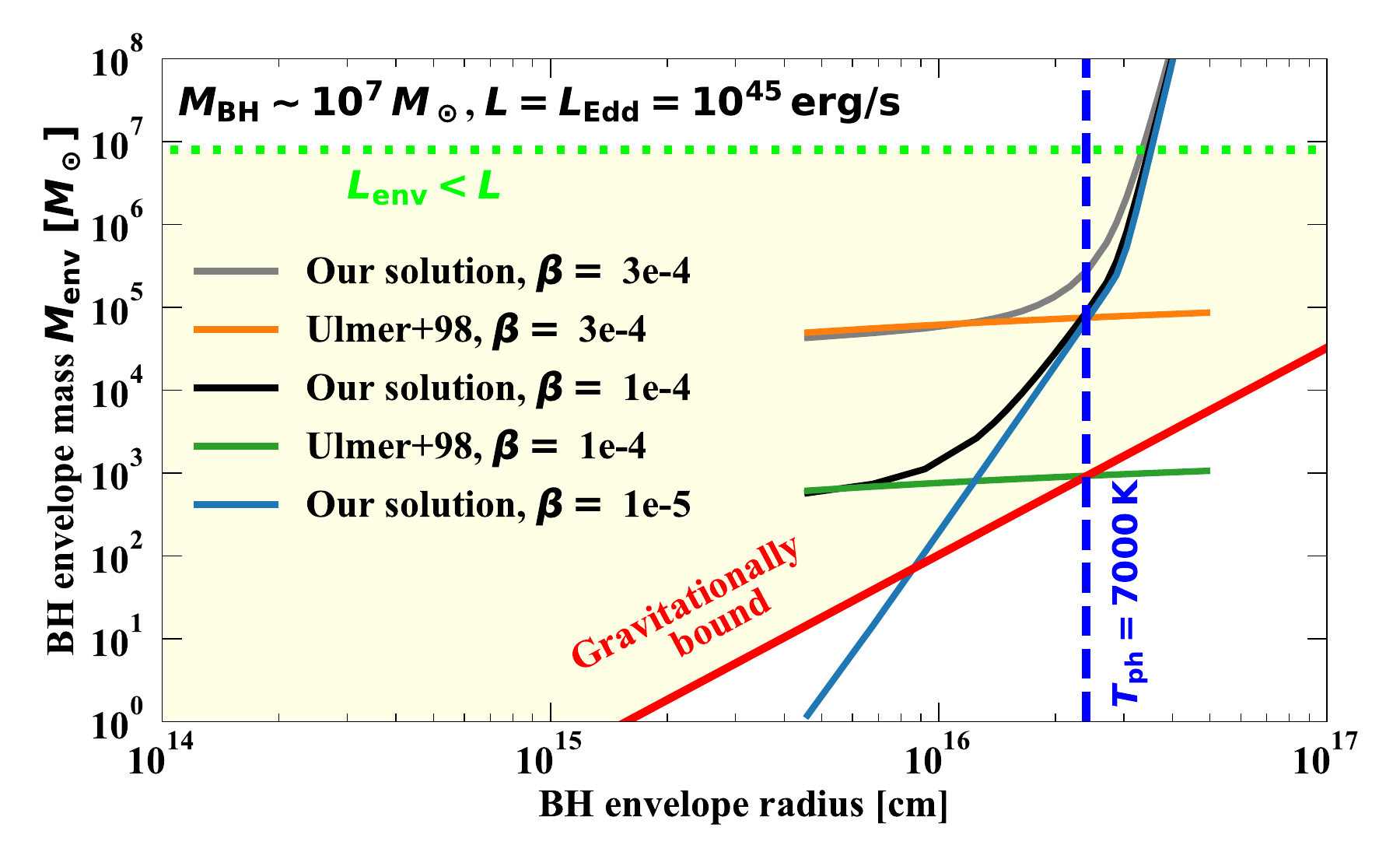}
    \caption{
      Minimum envelope mass required to be gravitationally bound for an Eddington luminosity $L=L_{\rm Edd}=10^{45}~\mathrm{erg~s^{-1}}$,
      as a function of the envelope radius (red solid line).
      The maximum envelope mass is also shown, for which
      its Eddington luminosity $L_{\rm env}$ is less than $L$ (green dotted line).
      The radius corresponding to a photospheric temperature of
      $T_{\rm ph}=7000~{\rm K}$ is also plotted (blue dashed line).
      We also present the envelope solution obtained in section \ref{sec:structure} and the solution derived in \citet{Ulmer1998A&A}. We adopt $\beta_c = 0.01$ here.
      Because of the beginning of hydrogen recombination, the solutions approach the same track, which resembles the behaviour of a massive star, the Hayashi track.
      Ulmer's solution corresponding to $\beta=10^{-5}$ is located entirely below $1M_\odot$ and is gravitationally unbound.}
    \label{fig:mass}
\end{figure}

The minimum mass of the BH envelope is constrained by the condition
that the envelope remains gravitationally bound to the BH
even after the outflow (and radiation) is injected over a dynamical time, 
$L r/v_{\rm esc}<M_{\rm env} v_{\rm esc}^{2}/2$,
which yields
\begin{align}
  M_{\rm env} &> M_{\rm env,min}=\frac{2 L r}{v_{\rm esc}^{3}},\nonumber\\
  &\sim 70 \left(\frac{L}{10^{45}~\mathrm{erg~s^{-1}}}\right)
  \left(\frac{M_{\rm BH}}{10^{7} M_{\odot}}\right)^{-3/2}
  \left(\frac{r}{10^{16}~{\rm cm}}\right)^{5/2} M_{\odot},
  \label{eq:Menv}
\end{align}
where the escape velocity is $v_{\rm esc}=\sqrt{2GM_{\rm BH}/r}$. 
This condition also implies that the luminosity $L$ must not exceed the maximum luminosity that can be carried by convection \citep{Begelman2008MNRAS,Coughlin2024ApJ}.
We plot this bound condition in Figure~\ref{fig:mass} assuming that the luminosity $L=10^{45}~\mathrm{erg~s^{-1}}$ is the Eddington luminosity of the BH, i.e., $M_{\rm BH}\sim 10^{7} M_{\odot}$.
The photospheric radius $r_{\rm ph}$ is 
\begin{align}
  r_{\rm ph} =
  \left(\frac{L}{4 \pi \sigma_{\rm sb} T_{\rm ph}^{4}}\right)^{1/2}
  \sim 2\times 10^{16} 
    \left(\frac{L}{10^{45}~\mathrm{erg~s^{-1}}}\right)^{1/2}
    \left(\frac{T_{\rm ph}}{7000\,{\rm K}}\right)^{-2}~\mathrm{cm},
    \label{eq:radius}
\end{align}
if the photospheric temperature of the envelope is $T_{\rm ph}$
where $\sigma_{\rm sb}$ is the Stefan–Boltzmann constant.
As we will discuss later, the photospheric temperature could be
$T_{\rm ph} \sim 7000~{\rm K}$,
which implies a radius of $r\sim 2\times 10^{16}~{\rm cm}$.
In this case, the envelope has $M_{\rm env} > 10^{3} M_{\odot}$
and is indeed optically thick.
The maximum envelope mass is determined by the condition that its Eddington luminosity $L_{\rm env} \equiv L_{\rm Edd}(M=M_{\rm env})$ does not exceed the observed luminosity $\sim L$, since a supermassive star typically radiates at the Eddtington limit.
We assume that the luminosity is primarily produced by accretion on to the BH, rather than by the envelope itself (i.e., through its gravitational contraction or nuclear burning), since the latter scenario corresponds to a supermassive star, which becomes unstable to general relativistic effects at envelope masses of 
$M_{\rm env} \sim 10^{7} M_{\odot}$ \citep{Iben1963ApJ, Chandrasekhar1964ApJ, Fowler1966ApJ}.
Note that the envelope mass can be smaller\footnote{
In the context of quasi-stars,
\cite{Ball+TZE:2011,Ball+TZ:2012} showed that
the BH mass is bounded by $M_{\rm BH} < 0.017 (M_{\rm env}+M_{\rm BH})$
by imposing an inner boundary condition at the Bondi radius,
and \cite{Coughlin2024ApJ} showed $M_{\rm BH} < 0.62 (M_{\rm env}+M_{\rm BH})$
by replacing the Bondi accretion with a saturated-convection region.
Since the central boundary condition remains uncertain
(due to the presence of a disc and outflows), we consider a broader parameter space, allowing for larger BH masses
(see also Sec.~\ref{sec:structure}).
} ($M_{\rm env}<M_{\rm BH}$)
or larger ($M_{\rm env} \ge M_{\rm BH}$) than the BH mass.

Mass infall from the ISM is indispensable for the envelope to survive over the duration of LRD activity
$t_{\rm out}$ in equation (\ref{eq:tout}),
especially when the envelope mass is smaller than the BH mass,
$M_{\rm env} < M_{\rm BH}$.
This is because the BH accretes at least at the Eddington rate
in equation (\ref{eq:dotMedd}),\footnote{
We also expect a wind from the envelope, but it could fall back again due to the recombination of hydrogen
\citep{Nakauchi+HOSN:2017}.
}
so that the envelope would be swallowed within a time-scale of
\begin{align}
  t_{\rm acc}=\frac{M_{\rm env}}{\dot{M}_{\rm Edd}}
  \sim 4.5 \times 10^{7} \left(\frac{M_{\rm env}}{M_{\rm BH}}\right)\,{\rm yr}, \label{eq:tacc}
\end{align}
which is shorter than $t_{\rm out}$ if $M_{\rm env} < M_{\rm BH}$.
Once the envelope mass decreases to the minimum mass in equation (\ref{eq:Menv}), it is blown away by the BH outflow.
To maintain a steady state, the mass infall rate from the ISM must balance the BH accretion rate
$\dot{M}_{\rm ISM} \sim \dot{M}_{\rm BH}$.
The mass infall rate from the ISM is determined by the gravitational collapse of the cloud. It may be roughly estimated as the Jeans
mass divided by the free-fall time,
\begin{align}
    \dot{M}_\mathrm{ISM} \sim \frac{c_s^3}{G} \sim 0.2 M_\odot\left( \frac{T}{10^4\,\mathrm{K}} \right)^{3/2}\, \mathrm{yr^{-1}}, \label{eq:dotMISM}
\end{align}
where $T$ is the ISM temperature. 
Stellar infall may also occur. The details of the accretion on to the BH envelope are highly uncertain, but as long as $\dot{M}_{\rm ISM} \gtrsim \dot{M}_\mathrm{BH},$ the BH envelope can grow.

The ISM infall would not be significantly hindered by emission from the envelope as long as the photospheric temperature remains below
$\sim 10^{4}~{\rm K}$, i.e.,
when the envelope is bloated \citep{Hosokawa+OY:2012,Hosokawa+YIOY:2013}.
For the mass infall rate in equation (\ref{eq:dotMISM}),
the envelope can continue to grow ($\dot{M}_{\rm ISM} \gtrsim \dot{M}_{\rm Edd}$) until the BH reaches a mass of $M_{\rm BH}\sim 10^{7} M_{\odot}$, assuming that the BH accretion is regulated at the Eddington limit.
Once the BH mass exceeds $M_{\rm BH} \gtrsim 10^{7} M_{\odot}$,
the envelope mass begins to decrease ($\dot{M}_{\rm ISM} \lesssim \dot{M}_{\rm Edd}$)
and is dispersed over the accretion time-scale $t_{\rm acc}$, marking the transition from an LRD to a normal AGN. This implies that the lifetime of the LRD is approximately equal to the Salpeter time $t_{\rm Sal}$ in equation (6) for a constant $\dot{M}_{\rm ISM}$.

The radius of the BH envelope is likely bloated to the photospheric radius $r_{\rm ph}$ in equation (\ref{eq:radius}) with a photospheric temperature $T_{\rm ph} \lesssim 10^{4}~{\rm K}$ under the ISM mass inflow,
as shown in the case of supermassive stars
\citep{Hosokawa+OY:2012,Hosokawa+YIOY:2013}.
As the total mass increases, the interior of the envelope contracts, but the gravitational energy released in the contraction is transported outward,
causing the surface layers to expand.
Alternatively, if the mass inflow rate is high,
the surface layers can also expand due to the deposition of gravitational energy near the surface.
The expansion time-scale is the Kelvin-Helmholtz time-scale in the former case,
and the mass inflow time-scale in the latter.
This expansion halts once the surface temperature drops below
$T_{\rm ph}\lesssim 10^{4}~{\rm K}$,
where the opacity sharply decreases.
Even in the absence of the ISM mass inflow, a wind from the envelope is also expected, since the surface opacity is approximately
$\kappa \sim \kappa_{\rm T}$
at $T_{\rm ph} \gtrsim 10^{4}~{\rm K}$
allowing the Eddington luminosity to blow off the surface layers.
The wind mass-loss rate is estimated as
\begin{align}
  \dot{M}_{\rm w} \sim 4\pi r^{2} \rho v_{\rm esc}
  \sim 0.3 \left(\frac{r}{3\times 10^{16}~\mathrm{cm}}\right)^{1/2}
  \left(\frac{M_{\rm BH}}{10^{7} M_{\odot}}\right)^{1/2} M_{\odot}~\mathrm{yr^{-1}}, \label{eq:dotMw}
\end{align}
where the surface density is estimated from the condition $\tau=\rho \kappa_{\rm T} r \sim 1$. Therefore, a significant amount of mass can escape from the surface as a wind if the surface temperature is higher than $T_{\rm ph}\gtrsim 10^{4}~{\rm K}$.
Once the temperature drops below $\sim 10^{4}~{\rm K}$ and the opacity decreases, the wind is expected to fall back \citep{Nakauchi+HOSN:2017}.
As a result, the envelope surface is likely to expand to
the photospheric radius $r_{\rm ph}$ in equation (\ref{eq:radius}) with a photospheric temperature $T_{\rm ph} \lesssim 10^{4}~{\rm K}$.
The dynamical time-scale near the surface is about
\begin{align}
    t_{\rm ph} \sim \frac{r_{\rm ph}}{v_{\rm esc}}
    \sim 
    2
    \left(\frac{L}{10^{45}~\mathrm{erg~s^{-1}}}\right)^{3/4}
    \left(\frac{T_{\rm ph}}{7000~{\rm K}}\right)^{-3}
    \left(\frac{M_{\rm BH}}{10^{7}M_{\odot}}\right)^{-1/2}~\rm{yr}. \label{eq:tph}
\end{align}
Remarkably, this is comparable to the observed variability time-scale \citep{Ji2025arXiv,Furtak2025A&A,D'Eugenio2025arXiv}.

Although we will not discuss the formation of the envelope in detail,
in order to form an envelope around a naked BH through mass infall from the ISM, the mass infall rate from the ISM must be significantly larger than
the BH's Eddington rate $\dot{M}_{\rm ISM} \gtrsim 100 \dot{M}_{\rm Edd}$
\citep{Inayoshi2016MNRAS}.
Otherwise, radiative and mechanical feedback from the BH will
suppress the infall, and the envelope mass will not reach the minimum mass
in equation (\ref{eq:Menv}).
For example, the inflow momentum is
$\dot{P}_{\rm in} \sim v_{\rm ff} \dot{M}_{\rm ISM} \sim 3
(v_{\rm ff}/10^{3}~\mathrm{km~s^{-1}})(\dot{M}_{\rm ISM}/M_{\odot}~\mathrm{yr^{-1}})
M_{\odot}~\mathrm{cm~s^{-2}}$
for the free-fall velocity $v_{\rm ff}$, which would not be sufficient to overcome the outflow momentum $\dot{P}_{\rm out}$ in equation (\ref{eq:dotPout}) if $\dot{M}_{\rm ISM} \sim \dot{M}_{\rm Edd}$.
When the inflow rate is sufficiently high,
$\dot{M}_{\rm ISM} \gtrsim 100 \dot{M}_{\rm Edd}$, the Bondi-like ISM inflow becomes optically thick at some point,
and photons begin to advect inward in the innermost region -- a process known as photon trapping
\citep{Inayoshi2016MNRAS}.
In this regime, the radiation feedback is no longer able to halt the Bondi inflow.
As the BH accretion rate is smaller than the inflow rate
($\dot{M}\propto r^{0.6}$ in the super-Eddington accretion discs)
mass accumulates, and an envelope begins to form.
The envelope mass continues to grow via the ISM inflow.
The envelope radius in equation~(\ref{eq:radius}) expands either due to the energy deposited by the inflow or
the luminosity from the central contraction
\citep{Hosokawa+OY:2012,Hosokawa+YIOY:2013}.
The expansion ceases once the surface temperature drops low enough for hydrogen recombination to occur, causing a sharp decline in opacity.
At this point, radiative feedback becomes ineffective
\citep{Hosokawa+YIOY:2013},
allowing even moderate inflow rates $\dot{M}_{\rm ISM} \gtrsim \dot{M}_{\rm Edd}$
to feed the envelope.
Thus we speculate that the existence of a hyper-accretion phase, $\dot{M}_{\rm ISM} \gtrsim 100 \dot{M}_{\rm Edd}$, during the early evolution may separate normal AGNs and LRDs, although further studies are needed, for example, on the effects of angular momentum and the dependence on BH mass.

A BH envelope could also form in direct-collapse models.
As a result of the collapse of a supermassive star,
a seed BH may form at the centre of a gaseous envelope.
This scenario has been discussed in the context of quasi-stars
\citep{Begelman2006MNRAS,Begelman2008MNRAS}.

\section{Structure and Emission of Envelope}\label{sec:structure}

A peculiar feature of LRDs is a red bump peaking at $\sim 0.5$--$1\,{\rm \mu m}$ in the rest-frame. This continuum flux may be described well with the thermal emission of a BH envelope with a luminosity $\sim L_{\rm Edd}$.
The colour of the BH envelope is determined by its structure. In the case of quasi-stars, $M_{\rm env}\gg M_{\rm BH}$, the surface temperature is $5000$--$7000\,{\rm K}$ \citep{Begelman2008MNRAS,Ball+TZE:2011}. 
This property results from the sharp decrease in the opacity at temperatures $T<10^{4}\,{\rm K}$ because of hydrogen recombination, analogous to the Hayashi track \citep{Hayashi1961PASJ, HayashiHoshi1961PASJ}.  
In this work, we focus on the other regime, $M_{\rm env, min}<M_{\rm env}\ll M_{\rm BH}$, and show that a similar track exists if the envelope mass is sufficiently large that the hydrogen recombination occurs in the outer envelope.

Here, we consider a hydrostatic envelope supported 
by the energy injection from a central BH with a luminosity
\begin{align}
    L = (1-\beta)L_{\rm Edd}, 
    \label{eq:L}
\end{align}
from an inner boundary radius $r_b\approx 10r_{\rm sch}$, where $P_\mathrm{gas}$ indicates the gas pressure at the photosphere, 
and $r_{\rm sch}$ is the Schwarzschild radius of the BH (see \citealt{Ulmer1998A&A}).
\dk{Importantly, $\beta$ appearing in eq. (\ref{eq:L}) is different from the gas fraction $p_\mathrm{gas} / p$.}
The hydrostatic structure of the envelope is computed with the toy opacity model in (\citealt{Begelman2008MNRAS}, see also Appendix \ref{ap:st}). This opacity model approximately captures the sharp decrease in the opacity with decreasing temperature due to hydrogen recombination.

In the case where the photospheric temperature is $T_{\rm ph}>10^4\,{\rm K}$ and the opacity is $\kappa_{\rm T} = 0.4~\mathrm{cm^2\,g^{-1}}$ throughout the entire envelope, 
the structure is described by a polytrope with $n=3$. \cite{Ulmer1998A&A} discuss the analytic solution in this regime and show that the envelope mass is very sensitive to $\beta$ as
\begin{align}
    M_{\rm env} \approx 1.6 \times 10^2 M_{\odot}\, \left(\frac{M_{\rm BH}}{10^{7}M_{\odot}}\right)^3
    \left(\frac{\beta}{10^{-4}}\right)^4
    \left[\ln\left(\frac{r_{\rm ph}}{r_b}\right)-1.8\right].
\end{align}
Note that the envelope mass is also sensitive to a hydrogen mass fraction $X$, where we take $X = 0.74$ and \dk{$Z=0$. }
Because of the logarithmic dependence on $r_{\rm ph}$, there is no specific value of $T_{\rm ph}$ for a given $M_{\rm BH}$ and $M_{\rm env}$ as long as $T_{\rm ph}\gtrsim 10^4\,{\rm K}$. 
However, for a fixed $\beta$, increasing the envelope mass $M_{\rm env}$ causes the photospheric radius to expand, which in turn causes $T_{\rm ph}$ to decrease. 
Eventually, $T_{\rm ph}$ falls below $10^4~\mathrm{K}$, which marks the onset of hydrogen recombination. 
In this regime, 
the photospheric radius is determined by the temperature where the $\mathrm{H^-}$ opacity starts decreasing, i.e., $T\sim 5000$--$7000\,{\rm K}$. The actual photospheric radius and temperature are determined by the structure of the outer radiative envelope, where convection is unable to carry all the flux. Here, we set the maximum convection flux as $F_\mathrm{con, max} = \beta_c p c_s$,
where $\beta_c$, $p$, and $c_s$ are the convection efficiency, total pressure, and isothermal sound speed, respectively \citep{Begelman2008MNRAS,Ball+TZE:2011,
Coughlin2024ApJ}.

\begin{figure}
    \centering
    \includegraphics[width=1\linewidth]{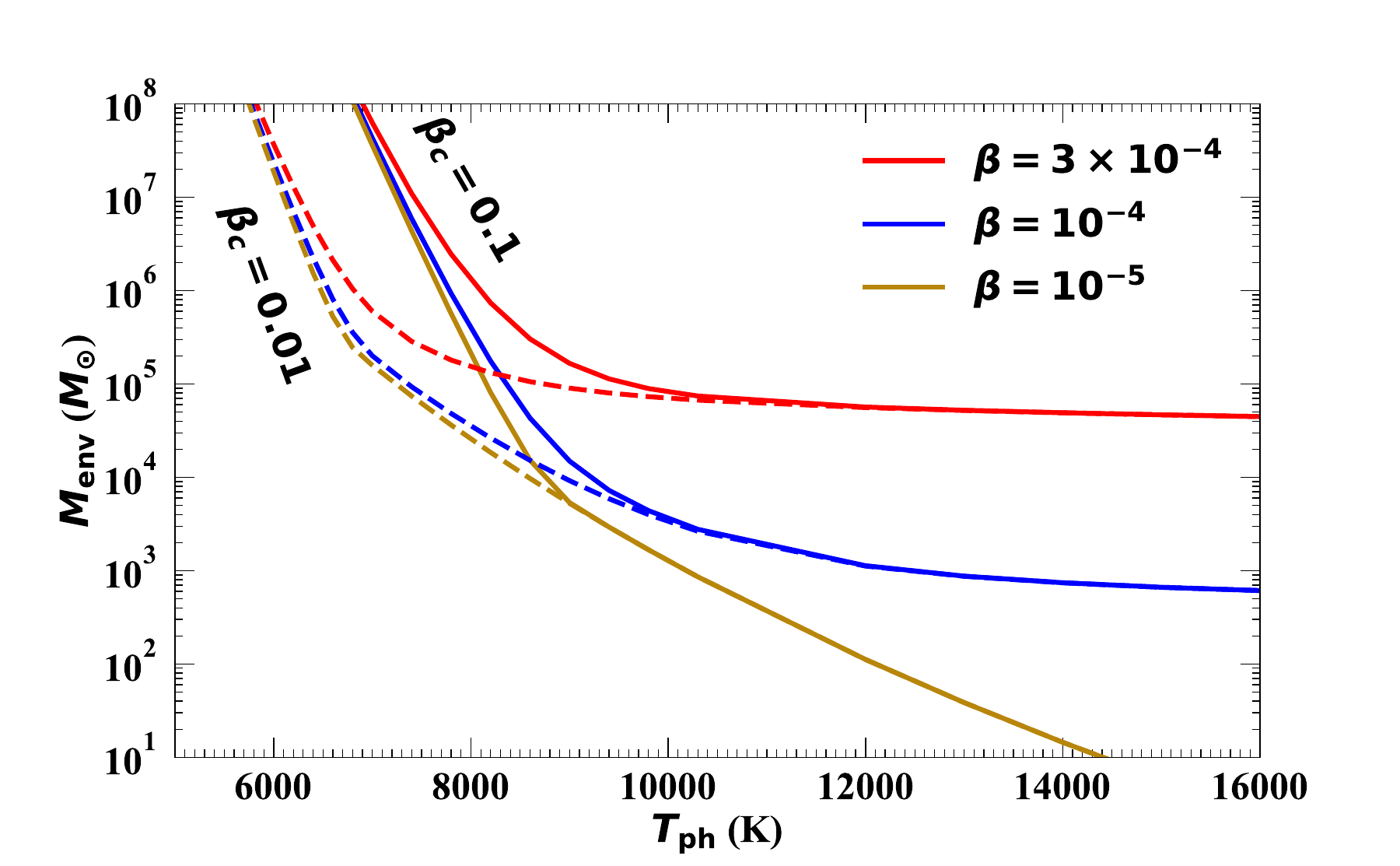}
    \caption{The envelope mass as a function of the effective temperature for a BH mass of $10^7M_{\odot}$. Solid and dashed curves correspond to the cases with a convection efficiency of $\beta_c=0.1$ and $0.01$, respectively. Here, the toy opacity model in \citet{Begelman2008MNRAS} is used (see equation \ref{eq:opc}).}
    \label{fig:M-T}
\end{figure}

Figure \ref{fig:M-T} shows the envelope mass as a function of $T_{\rm ph}$ for a given $\beta$ and $\beta_c$ in the case of $M_{\rm BH}=10^{7}M_{\odot}$. For $T_{\rm ph}<10^4\,{\rm K}$, the curves for a given $\beta_c$ converge into a single track. The photospheric temperature of the track is $7000$--$8000\,{\rm K}$ for $\beta_c=0.1$ and $6000$--$7000\,{\rm K}$ for $\beta_c=0.01$. 
The envelope mass at a given $T_{\rm ph}$
decreases with decreasing $\beta_c$ because the density inversion in the radiative envelope is more significant when the convection efficiency is lower.
Note that our result is likely to overestimate the photospheric temperature because the toy opacity model does not include the opacity bump due to bound-free and bound-bound absorption \citep[see, e.g.,][]{Begelman2006MNRAS}. Thus, we expect these red tracks to be shifted towards slightly lower temperatures if a realistic opacity table is used. 
\dk{In the present setup, we employ a 1D hydrostatic envelope. 
We emphasize, however, that this does not preclude real microturbulence in the envelope. Indeed, small‑scale turbulent motions may well be present. 
In fact, the density inversion near the surface in our model implies hydrodynamical instability. 
Furthermore, macroscopic kinematics, such as rotation, inflow, and outflow can produce comparable Doppler broadening of the Balmer limit. 
A full 3D radiative transfer treatment including both microturbulence and large‑scale velocity fields will be required to distinguish their relative contributions.}

In the regime where the ISM inflow on to the envelope is strong, so that the photosphere is located within the inflow instead of the envelope, the appearance of the envelope changes. In this case, the photosphere is determined by the condition $\kappa \rho r\sim 1$, and due to the onset of hydrogen recombination, the opacity plummets as temperature goes down. We follow the same argument presented in \cite{Stahler1986ApJ} but employ the toy opacity presented in equation (\ref{eq:opc}).
We estimate that the photospheric radius and the photospheric temperature
can be described as 
\begin{align}
r_\mathrm{ph} &= 2.0 \times 10^{16} ~ \left( \frac{\dot{M}_\mathrm{ISM} }{10\dot{M}_\mathrm{Edd}}\right)^{1/7}\left( \frac{M_\mathrm{BH}}{10^7M_\odot}\right)^{11/28}~\mathrm{cm} \label{eq:r_ph_stahler},  \\
T_\mathrm{ph} &= 7000 ~ \left( \frac{\dot{M}_\mathrm{ISM} }{10\dot{M}_\mathrm{Edd}}\right)^{-1/14} \left( \frac{M_\mathrm{BH}}{10^7M_\odot}\right)^{3/56} ~\mathrm{K}\label{eq:t_ph_stahler} ,
\end{align}
yielding similar results as the case where the photosphere is embedded within the envelope. 
We note that our scalings with $\dot{M}_\mathrm{ISM}$ and $M_\mathrm{BH}$ differ from those of \citet{Stahler1986ApJ}, since we adopt a constant Eddington luminosity whereas they determine $L$ self-consistently by solving the heat equation with $\dot{M}_\mathrm{ISM}$ dependent luminosity. Nevertheless, in both treatments the photospheric temperature remains close to $7000\,\mathrm{K}$ and exhibits only a weak dependence on the mass infall rate. 

\begin{figure}
    \centering
    \includegraphics[width=1\linewidth]{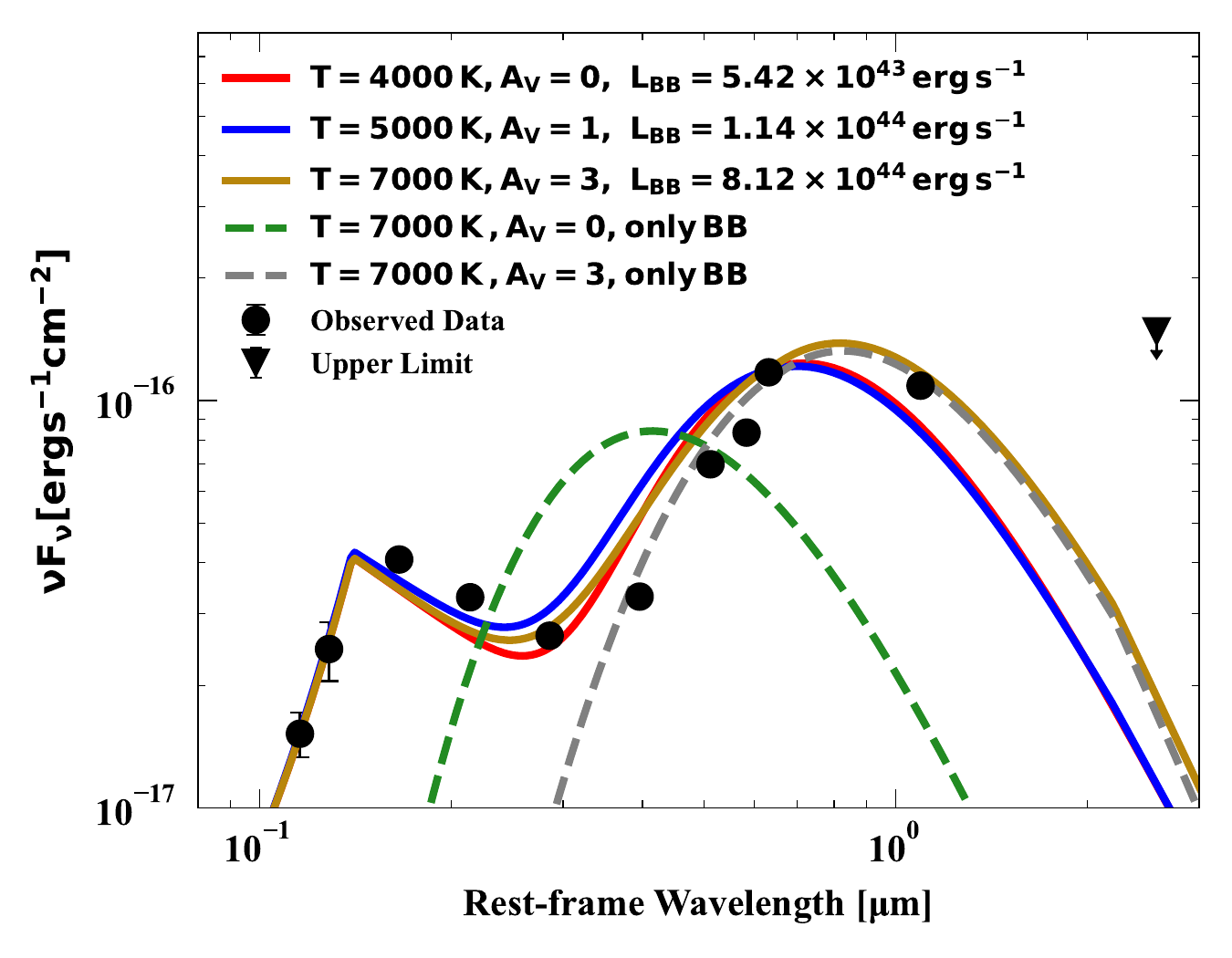}
    \caption{
    SED of stacked LRDs with theoretical curves obtained by our model. Stacked SED data points are taken from Table 4 in \citet{Akins2024arXiv}.
    Each line shows the envelopes with different photospheric temperatures and $A_V$. 
    \dk{The green and grey dashed line highlights the blackbody component with $T = 7000~\mathrm{K}$ and $A_V = 0, 3$ respectively.}
    We adopted a simple SED model that includes the star-formation effect as a power law and set the Lyman break around $0.1$ micron artificially since our work focuses on the redder part of the SED. The dust attenuations are included by the Calzetti dust extinction law \citep{Calzetti2000ApJ}.}
    \label{fig:SEDfit}
\end{figure}

Figure \ref{fig:SEDfit} compares the stacked spectral energy distribution (SED)
provided by \cite{Akins2024arXiv} with model SEDs computed as the sum of blackbody emission from a BH envelope and emission from star formation. 
As they utilize both the COSMOS-Web and the PRIMER-COSMOS survey data, the sample size ranges from 29 to 434 objects across the different bands.
For the emission from the BH envelope, we estimate the luminosity by fitting blackbody models
assuming three parameter combinations -- $(T_{\rm ph},~A_V)= ( 4000\,{\rm K}, ~0)$, $(5000\,{\rm K}, 1)$, and $(7000\,{\rm K}, ~3)$ -- where the attenuation law is adopted from \citet{Calzetti2000ApJ}.
In addition, we add the emission from star formation to each of the model spectra, assuming a simple power law, which dominates the bluer part of the SED. 
The photospheric temperature obtained by our hydrostatic model gives SEDs consistent with those observed in LRDs. 
The fit also gives luminosities in agreement with our model, provided that the central BH mass is on the order of $ 10^7 M_\odot$. 
\dk{
Our model with $T_\mathrm{ph} = 7000\,\mathrm{K}$ and no extinction shows clear tension with the stacked photometric data. Given recent studies suggesting a dust deficit in the LRD environment \citep{Setton2025arXiv, Akins2025arXiv, Xiao2025arXiv}, the allowed temperature range becomes more constrained. We therefore conclude that reproducing the observed SEDs requires BH envelope temperature in the range $4000\text{--}5000~\mathrm{K}$ under such a dust-poor environment.
}
Note that we assumed the star formation emission is not attenuated and the redshift of the stacked data is \dk{6}, but the SED actually consists of stacked data of $z=4\text{--}7$, with each individual LRD having a distinct redshift and $A_V$. 
\section{Conclusions}
\label{sec:conclusions}
Unprecedented observations by the JWST have revealed a new type of AGNs, so-called Little Red Dots (LRDs), that exhibit a unique V-shaped SEDs with a red optical continuum and an excess at UV bands, while showing broad (FWHM $\gtrsim1000\,\mathrm{km~s^{-1}}$) hydrogen Balmer emission lines that indicate the presence of accreting SMBHs.
In this study, we show that in a standard super-Eddington disc scenario, momentum feedback from the strong outflows, assuming conservative values of the duty cycle ($f_\mathrm{duty} = 0.03$) and outflow efficiency ($\xi = 0.03$), would inject sufficient momentum to make the entire gas gravitationally unbound as shown in Figure \ref{fig:feedback}. 
We stress that, despite its reliance on highly idealized assumptions, this argument produces robust results and may therefore be applied to other galaxies hosting massive central BHs. 
In addition to momentum feedback, the outflow likely would disrupt the LRD structure through feedback due to energy deposition or otherwise generate X-ray emission, potentially in conflict with current X-ray observations.
To resolve this apparent discrepancy, we propose the existence of a dense, extended envelope around the central BH that tightly confines the outflow and reprocesses its energy into blackbody radiation from the envelope's surface. 
A lower limit on the envelope mass follows directly from requiring that the envelope remains gravitationally bound. 
By imposing hydrostatic equilibrium, we demonstrate that such an envelope with a photospheric radius around $3\times 10^{16}~\mathrm{cm}$ naturally attains a photospheric temperature $5000\sim 7000~\mathrm{K}$ owing to the steep drop in opacity at the onset of the hydrogen recombination. 

The BH envelope can intrinsically suppress the feedback from the outflow and produce the red continuum widely observed in LRDs. 
The dynamical time-scale of the envelope naturally matches the observed variability time-scale of longer than a year. 
The BH envelope can grow as long as the infall rate at its surface satisfies $\dot{M}_\mathrm{ISM} \gtrsim \dot{M}_\mathrm{Edd}$. 
If the BH mass grows or the mass infall decreases (i.e., $\dot{M}_\mathrm{ISM} \lesssim \dot{M}_\mathrm{Edd}$),
the BH envelope loses mass and disperses, trasforming the LRD back into a normal AGN over a Salpeter time.
Such an envelope could naturally develop around the central BH as a result of super-Eddington accretion $\dot{M}_\mathrm{ISM} \gtrsim100\dot{M}_\mathrm{Edd}$, at least during the early phase of the formation. 

While the BH envelope model reproduces the main spectral features of LRDs and suppresses the outflow, it leaves several key questions unanswered, such as the nature of the UV ionizing source and the origin of the broad-line region. 
Recent work incorporating additional physics suggests that electron scattering may play a role in broadening the emission lines, leading to a narrower intrinsic FWHM for the broad line region than previously thought \citep{Weymann1970ApJ, Huang2018MNRAS, Rusakov2025arXiv}.
Physical processes on the envelope surface, such as accretion shocks, dynamical instabilities, and magnetic reconnection, may also affect the line profiles and act as a UV ionization source \citep[\dk{see e.g. Figure 1 in}][]{Rodriguez2025PASP}. 
Likewise, the formation and evolution of the envelope, including the effects of asphericity \dk{such as the collimation of the outflow \citep{King2025MNRAS}}, also deserve further study in future work. 
Nevertheless, the model presented here offers important insights into the growth of SMBHs in the early universe and the origin of LRDs.

\section*{Acknowledgements}
We thank Masaru Shibata, Takashi Hosokawa, Kazumi Kashiyama, Kazuyuki Omukai, Naoki Yoshida, \dk{Yuto Ogasawara,} Yurina Nakazato, Yuichi Harikane, and Darach Watson
for helpful comments and discussions. 
The work of D. Kido is supported by the Forefront Physics and Mathematics Program to Drive Transformation (FoPM).
K. Ioka acknowledges support from KAKENHI 22H00130, 23H05430, 23H04900, 23H01172.
K. Hotokezaka and C M. Irwin are supported by the JST FOREST Program (JPMJFR2136) and the JSPS Grant-in-Aid for Scientific Research (20H05639, 20H00158, 23H01169, 23H04900). 
K. Inayoshi acknowledges support from the National Natural Science Foundation of China (12233001), 
the National Key R\&D Program of China (2022YFF0503401), and the China Manned Space Project (CMS-CSST-2025-A09). 
\dk{We thank the anonymous referee for the useful comments on this manuscript.}

 \section*{Data Availability}
The data that support the findings of this study are available from the corresponding author upon reasonable request.



\bibliographystyle{mnras}
\bibliography{ref} 




\appendix

\section{BH envelope Structure}\label{ap:st}
Let us consider the structure of a hydrostatic envelope
surrounding a BH with a mass, $M_{\rm BH}$. Because the details of the vicinity of the BH are uncertain,
we consider the situation where the envelope is supported by the energy injection from an accreting BH with luminosity, $L=(1-\beta)L_{\rm Edd}$, at an inner boundary $r_b$ \citep{Loeb1997ApJ, Ulmer1998A&A}.
The envelope structure is described by the equation of hydrostatic equilibrium,
\begin{align}
    \frac{dp}{dr} &= -\frac{GM_{\rm BH}}{r^2}\rho,\label{eq:hydro}
\end{align}
the equation of state,
\begin{align}
    p=p_g+p_r=\frac{\rho k_BT}{\mu m_h} + \frac{a}{3}T^4, \label{eq:eos}
\end{align}
and the energy conservation (see \citet{Kippenhahn1990}), 
\begin{align}
    \frac{dT_{\rm rad}}{dr} & = -\frac{3}{4ac}\frac{L}{4\pi r^2}\frac{\kappa \rho}{T^3},\\
    \frac{dT_{\rm ad}}{dr} & = \frac{8-6(p_g/p)}{32-24(p_g/p)-3(p_g/p)^2}\frac{T}{p}\frac{dp}{dr}.\label{eq:Tad}
\end{align}
The actual temperature gradient is given by $dT/dr={\rm max}(dT_{\rm rad}/dr,dT_{\rm ad}/dr)$.
As discussed by \cite{Begelman2008MNRAS}, if the surface temperature is low ($<10^4\,{\rm K}$), it is important to note that there exists an outer radiative zone since the convection is unable to carry all the energy flux. To take this effect into account, we set the maximum convection flux as $F_\mathrm{con, max} = \beta_c p c_s$, where $\beta_c$ is the convection efficiency and $c_s=\sqrt{p/\rho}$ is the local isothermal sound speed. 

Numerically, we set the temperature gradient as 
\begin{align}
\frac{dT}{dr} = \frac{dT_{\rm rad}}{dr} - \frac{\left[ \mathrm{min} \left( - \frac{dT_\mathrm{rad}}{dr}, - \frac{dT_\mathrm{ad}}{dr}\right)  + \frac{dT_\mathrm{rad}}{dr}\right]}{\left( 1 + x^{10} \right)} \label{eq:dtdr}
\end{align}
\dk{where $x = F / F_\mathrm{con, max}$ and $F = L /  4\pi r^2$ is the total energy flux.}
Here, we adopt the toy opacity proposed by \cite{Begelman2008MNRAS}:
\begin{align}
    \kappa(T) \approx \frac{\kappa_{\rm T}}{1+(T/T_0)^{-s}},\label{eq:opc}
\end{align}
where $s=13$ and $T_0=8000\,{\rm K}$.
This toy model approximately 
captures the sharp decrease in the opacity at $T\lesssim 10^4\,{\rm K}$.

We numerically integrate equations (\ref{eq:hydro})--(\ref{eq:Tad}) inward from the surface for various values of $r_\mathrm{ph}$, thereby giving the corresponding $T_\mathrm{ph}$ from $L$ and $r_\mathrm{ph}$.
To obtain a unique solution for the equations (\ref{eq:hydro}) and (\ref{eq:dtdr}), an additional boundary condition at the surface is necessary. We adopt the same boundary condition as in \cite{Begelman2008MNRAS} at the photosphere $r_{\rm ph}$:
\begin{align}
    p(r_{\rm ph}) = \frac{2}{3}\frac{g}{\kappa(T_{\rm ph})}+\frac{aT_\mathrm{ph}^4}{6},
\end{align}
where $g=GM_{\rm BH}/r_{\rm ph}^2$ is the surface gravity and $T_{\rm ph}$ is the photospheric temperature.
This condition ensures that the optical depth at the photosphere is $2/3$ and properly accounts for both gas and radiation pressure contributions. Note that the factor $1/6$ is due to the outward contribution of the radiation pressure.
Finally, the density at the surface can be obtained by the equation of state (\ref{eq:eos}).

\section{Notations}\label{ap:not}
\begin{table}
    \centering
    \begin{tabular}{ccc}
        notation & Explanation & Equations \\ \hline
         & Rate notations & \\
        $\dot{M}_\mathrm{Edd}$ & Eddington accretion rate & \ref{eq:dotMedd} \\
        $\dot{M}_\mathrm{ISM}$ & Mass infall rate from the ISM on to the envelope & \ref{eq:dotMISM} \\
        $\dot{M}_\mathrm{w}$ &  Wind mass-loss rate & \ref{eq:dotMw} \\
        $\dot{m}$ & Normalized BH accretion rate &  \ref{eq:dotm}\\
        $\dot{P}_\mathrm{out}$ & Momentum outflow rate & \ref{eq:dotPout} \\
        \hline
         & Mass notations & \\
         $M_\mathrm{BH}$ & Black hole mass & \\
         $M_\mathrm{gas}$ & Gas mass in LRD & \\
         $M_*$ & Stellar mass in LRD & \\
         $M_\mathrm{env}$ & Mass of the envelope & \ref{eq:Menv} \\
         $M_\mathrm{env, min}$ & Minimum Mass of the envelope & \ref{eq:Menv} \\ \hline 
          & Time-scale notations &  \\
          $t_\mathrm{out}$ & Total duration of the outflow & \ref{eq:tout} \\
          $t_\mathrm{Sal}$ & Salpeter time & \ref{eq:tSal} \\
          $t_\mathrm{age}$ & the cosmic age of the universe at a given redshift  & \\
          $t_\mathrm{acc}$ & Accretion time of the envelope to the BH  & \ref{eq:tacc} \\
          $t_\mathrm{ph}$ & Dynamical time-scale near the surface of the envelope & \ref{eq:tph} \\ \hline 
           & Length notations & \\
           $r_\mathrm{LRD}$ & Typical radius of LRDs observed so far & \\
           $r_\mathrm{ph}$ & Photospheric radius of the envelope & \ref{eq:radius}, \ref{eq:r_ph_stahler} \\
           $r_b$ & Inner boundary radius of the envelope & \\ \hline
            & Other notations & \\
            $L$ & Luminosity of the envelope & \ref{eq:L}\\
            $L_\mathrm{Edd}$ & Eddington luminosity & \\
            $\beta$ & Parameter that determines the fixed luminosity & \ref{eq:L} \\
            $\beta_c$ & Convection efficiency &  \\
            $f_\mathrm{duty}$ & Duty cycle of the AGN activity & \\
            $P_\mathrm{out}$ & Total momentum produced by the outflow within its duration & \ref{eq:Pout} \\
            $P_\mathrm{LRD}$ & Total required momentum to make LRD unbound & \ref{eq:PLRD} \\
            $T_\mathrm{ph}$ & Photospheric temperature & \ref{eq:t_ph_stahler} 
    \end{tabular}
    \caption{Notations throughout the paper.}
    \label{tab:not}
\end{table}
\bsp	
\label{lastpage}
\end{document}